\RequirePackage{lineno}
\documentclass[%
 reprint,
twocolumn,
showpacs,preprintnumbers,showkeys,
 amsmath,amssymb,
 aps,
prc,
floatfix,
]{revtex4}

\usepackage{graphicx}% Include figure files
\usepackage{dcolumn}% Align table columns on decimal point
\usepackage{bm}% bold math
\usepackage{amssymb}
\usepackage{float}
\usepackage[caption = false]{subfig}
\usepackage{hyperref}% add hypertext capabiliti
\begin{document}

%\preprint{APS/123-QED}

\title{Testing of coalescence mechanism in high energy heavy ion collisions using two-particle correlations with identified particle trigger}
\author{Subikash Choudhury}
\email{subikash.choudhury@cern.ch}
\author{Debojit Sarkar}
\author{Subhasis Chattopadhyay}
\email{sub@vecc.gov.in}
\affiliation{Variable Energy Cyclotron Centre, Kolkata India}
\date{\today}

%\corauth[cor]{Corresponding author.}
%\ead{sub.chattopadhyay@gmail.com}
%\address[label1]{Variable Energy Cyclotron Centre, 1/AF-Bidhannagar, Kolkata-700064, India}

\begin{abstract}
In central Au-Au collisions at top RHIC energy, two particle correlation measurements with identified hadron trigger 
have shown attenuation of near side proton triggered jet-like yield  at intermediate transverse momentum ($p{_T}$), 
2$< p{_T} <$ 6 GeV/$\it{c}$.
The attenuation has been attributed to the anomalous baryon enhancement observed in the
single inclusive measurements at the same $p{_T}$ range. The enhancement has been found to be in agreement with 
the models invoking coalescence of quarks as a mechanism of hadronization.
Baryon enhancement has also been observed at LHC in the single inclusive spectra.
We study the consequence of such an enhancement on two particle correlations at 
LHC energy within the framework of A Multi Phase Transport (AMPT) model that implements quark coalescence as a mode of hadronization.
In this paper we have calculated the proton over pion ratio and the near side per trigger yield associated to pion and proton
triggers at intermediate $p{_T}$ from String Melting (SM) version 
of AMPT.
Results obtained are contrasted with the AMPT Default (Def.)  which does not include coalescence.
Baryon enhancement has been observed in AMPT SM at intermediate $p{_T}$. 
Near side jet-like correlated yield associated to baryon (proton) trigger in the momentum region where baryon generation
is enhanced is found to be suppressed as compared to the corresponding yields for the meson (pion) trigger in most central
Pb-Pb events.
No such effect has been found in the Default version of AMPT.
\end{abstract}

%\item{PACS numbers}
%\begin{keyword}


%\end{keyword}
%\end{frontmatter}

\maketitle

\section{Introduction}

Temperature and energy density attained in ultrarelativistic heavy ion collisions
at RHIC and LHC are compatible with the lattice quantum chromodynamics (l-QCD) thresholds for phase transition 
from hadronic to a de-confined state of quarks and gluons \cite{PHCPRL104_2010, ALCNPA573c_2013, FKarNPA698_2002}.
The Hot and dense matter thus formed, known as quark-gluon plasma (QGP), cools down in the process of evolution, 
re-confines to hadrons and streams freely to detectors \cite{MKLNP785_2010}. This gives us a unique opportunity
to study the mechanism of particle production under extreme conditions.
The $p_T$ spectra of the final state particles give an in-sight on their
production mechanism and of interactions at various stages of evolution \cite{STCNPA715_2003, 
PHCPRC69_2004,ALCPRC88_2013}.
Various theoretical models have been proposed but no unique prescription is available to explain 
the $p{_T}$ spectra over the entire experimentally measured range.
Particle production below $p{_T} $ $\approx$ 2 GeV/$\it{c}$, referred to as the bulk region
can be reproduced by the hydro-inspired models \cite{SBPRC61_2000, DTPRL86_2001, THPRC77_2008}.
For $p{_T}  >$ 6  GeV/$\it{c}$, hadronization is primarily through fragmentation of high  $p{_T}$ partons to a collimated
shower of hadrons (jets). This process involves parton scattering with large momentum transfer and can be 
convincingly described by perturbative QCD calculations \cite{FAEPJC61_2009, DEntLNP785_2010}. 

However, none of these approaches could account for particle production at 
the intermediate  $p{_T}$  (2$<p{_T}<$6 GeV/$\it{c}$). The observations like the anomalous enhancement of inclusive  
baryon over meson yield
, particle species dependence of the nuclear modification factor (R$_{AA}$,R$_{CP}$)
 and baryon-meson ordering of elliptic flow coefficient ($v_{2}$) were found to be at odds with
either of these formalisms \cite{PHCPRC69_2004, PHCPRL98_2007}.
Plausible explanations to enhanced baryon/meson or nuclear modifications were achieved from the models 
either incorporating recombination of quarks \cite{RFPRL90_2003,RJFPRC68, VGPRL90_2003,RHwa_PRC70_2004} or boost from a 
radially expanding medium pushing massive hadrons to higher $p_{T}$ (Hydrodynamics) \cite{SBPRC61_2000, THPRC77_2008}. In principle, both the approaches
attempt to generate high $p_{T}$ baryons from soft processes as opposed to mesons.

Another explanation could be in terms of energy loss of partons in the medium.
The independent fragmentation of energetic partons based on pQCD calculations gives baryon/meson $\sim$
0.1 both in light and strange flavor sectors \cite{RJFPRC68}. This is in contradiction to the experimental results. 
However, jet fragmentations are strongly influenced by the dense medium leading to an alteration 
of the fragmentation function \cite{PHCPRL90_2003, ALCPRL108_2012}.
It has been argued that the medium modified fragmentation can also be a potential source of enhanced baryon generation
\cite{UAWEPJC55_2008}.
Jet-like peak structure observed in the correlation measurements between baryons and charged hadrons at intermediate $p{_T}$ 
reported by the PHENIX and STAR Collaborations may be an indication that the baryon enhancement is associated to the medium induced jet 
modification \cite{PHCPRC71_2004, STCarXiv1410_2014}.

The high density environment achieved in heavy-ion collisions 
may be conducive for hadron formation through coalescence of quarks. In simple coalescence approach, quark and anti-quark pairs close in phase space 
recombine into mesons and three quarks to (anti-)baryons. Thus at the same $p{_T}$, baryons and mesons are formed from the 
quarks with momenta $\sim p{_T}$/3 and $\sim p{_T}$/2 respectively. Different approaches of quark recombination have been 
suggested and adopted by various groups. Each of them particularly differ in the way high $p{_T}$ partons from the 
initial hard-scatterings and the thermalized soft partons are treated. While some consider coalescence of only soft partons and 
hard partons to hadronize by fragmentation only \cite{RFPRL90_2003, RJFPRC68}, others allow coalescence 
of both soft and hard minijet partons \cite{RHwa_PRC70_2004} . 
Since the p${_T}$ spectra of these  hard partons show a power-law behaviour, an exponential thermal spectrum of 
soft  partons is therefore imperative for large baryon to meson enhancement. All these implementations with proper 
tuning of parameters describe the basic features at intermediate p${_T}$ e.g., p${_T}$ spectra, $v_{2}$-scaling reasonably well at 
RHIC energy. At LHC, scaling violation of $v_{2}$ is somewhat larger than that at RHIC and may be naturally explained within soft-hard recombination 
formalism \cite{RHwa_PRC78_2008}. Additionally, the near-side peak structure observed in the measurements of azimuthal correlations
 triggered by identified particles at intermediate p$_{T}$ at RHIC energy
have been reasonably explained with the inclusion of mini-jet partons or partons from hard scatterings in the coalescence formalism. 
Thus an alternative way to look for the source of baryon anomaly at intermediate p$_{T}$
 is to study the baryon-charged hadron correlations.
The angular correlation measurements are likely to be more sensitive to probe the contribution
of hard scattering towards hadron production. In this paper we have studied the sensitivity of di-hadron correlation 
measurements to the coalescence mechanisms when measured by itaking dentified mesons ($\pi$) and baryons ($\it{p/\bar{p}}$) at 
intermediate p${_T}$ as leading hadrons.

The two-particle azimuthal correlation functions triggered by leading hadrons encode the characteristic of 
the production mechanism of the trigger and associated particles. The correlation measurements with 
high $p_{T}$ trigger particles ($>$ 4-6 GeV/$\it{c}$) in p-p collisions manifest itself as di-jet peaks in azimuth, an imprint of the
QCD fragmentation of  back to back hard scattered  partons \cite{STCPRL95_2005}. At intermediate $p_{T}$, 
hadronization through recombination would lead to ``trigger dilution'' in central heavy ion collisions
\cite{PHCPRC71_2004, STCarXiv1410_2014}.
Trigger particles originating from recombination or coalescence of thermal quarks 
from the dense partonic medium would lack correlated hadrons at small angular region (jet-like correlation).
This would effectively dilute (reduce) per trigger associated yield. Furthermore, 
dilution is expected to be prominent for baryon trigger than meson trigger as the baryon production is more favourable 
through coalescence of quarks. 
%The ratio of the yields in the near side peak for baryon and meson triggered correlations 
%can therefore be an important tool to investigate the role of soft and hard partons in the process of recombination. 
 
In this work, the sensitivity of the near side yields of proton and pion triggered azimuthal 
correlation functions to the coalescence mechanism have been tested using two versions of the AMPT model
\cite{ZWLPRC72_2005}. 
While the partonic version (SM) of the AMPT model produces particles by the coalescence of quarks, 
the default version has only minijets and strings fragmenting to hadrons.
We have built triggered correlation functions from the events generated  from  either version of 
the AMPT model for Pb-Pb collision at $\sqrt{s_{NN}}$ = 2.76 TeV and extracted near side yield as final observable. 

The paper is organised as follows. 
In the next section we have given a brief introduction to the AMPT model and the implementation of coalescence mechanism. 
The method of extraction of background subtracted correlation function is discussed in section 3. 
The results and discussions are performed in sections 4 and 5 respectively.

\section{The AMPT Model}

The AMPT model has been extensively studied at RHIC and LHC energies. 
Free parameters of the model have been constrained by a wide range of experimental data. 
If broadly classified, model has two modes: Default ( minijets and strings ) and String Melting (replicate QGP  
allowing strings to melt into partons) \cite{ZWLPRC65}. The spatial and momentum distributions of minijet partons and 
excited soft strings, obtained from the HIJING model \cite{XNWPRD44} are used as initial conditions for subsequent modeling of partonic evolution. 
In the default version, minijet partons are evolved via a parton cascade model (ZPC)  \cite{BZCPC109} which basically includes 2-body elastic 
scatterings among the partons with a medium dependent scattering cross-section represented as ${\sigma_{p}\simeq 9\pi\alpha_{s}^{2}/2\mu^{2}}$ 
where $\alpha_{s}$ is the QCD coupling constant for strong interactions and $\mu$ is the Debye screening mass of gluons in QGP medium.
Although, it is a function of temperature and density of the partonic medium but in ZPC it is parameterised to fix 
the magnitude of scattering cross-section. At the end of evolution, these minijet partons are recombined with their parent strings
and are eventually hadronized  by the Lund string fragmentation \cite{TSCPC82}. The post hadronization stage is modeled by A Relativistic Transport 
model (ART) \cite{BALIPRC52, BALIIJPE10}, which guides the hadronic interactions till freeze-out.

To emulate the conditions similar to the de-confined QGP, AMPT has been extended to perform melting of excited strings. 
Taking  initial conditions from HIJING, strings are first fragmented to hadrons followed by conversion of these hadrons to 
valance quarks/antiquarks preserving their flavor and spin quanta. Now the system comprises both minijet and string melted 
partons, which are further scattered through ZPC. Once the interaction ceases, partons are re-confined to hardons via 
an implementation of coalescence formalism that combines two or three partons nearest in coordinate space to mesons and/or 
(anti-)baryons respectively. Mass and flavor of  hadrons are determined from the invariant mass and respective flavors of the
coalescing partons. Therefore a quark-antiquark pair will be recombined to pions provided di-quark invariant mass is in the proximity of pion mass. 
Present approach of coalescence is therefore not exactly similar to those discussed in \cite{RFPRL90_2003, VGPRL90_2003,RHwa_PRC70_2004}.
Here it allows coalescence of
partons with a relatively large momentum difference. To account for the hadronic interactions prior to freeze-out, final state hadrons are then 
transported through ART model.

AMPT in SM mode with partonic cross section of 6-10 mb provides a good fit to the flow observables at top RHIC energy \cite{SAVNPA715_2003}. 
While at LHC, with increased beam energy and high initial temperature, data seem to be better reproduced with
the choice of a lower parton scattering cross section \cite{SPALPLB709_2012}. In this study we have set scattering 
cross section to 1.5 mb by tuning ${\alpha_{s}}$= 0.33  and  ${\mu}$= 3.22 $ \it{fm^{-1}}$ keeping in mind that this 
particular choice simultaneously reproduces charged particle multiplicity density and flow coefficients at LHC energy 
\cite{ALCPLB_2013, JXu11012231v2}. The parameters for Lund string fragmentation are kept same as that of the default 
HIJING values corresponding to smaller string tension \cite{JXu11012231v2}.

\section{Analysis Method}
%{\bf this section needs to written in more detailed, it is not clear from the present draft how the whole analysis is done, please write step by step.}

%Two particle correlation is a widely used technique in heavy-ion physics for extracting the properties of the medium 
%produced in the collision. This technique also assists in understanding of the jet-medium interplay
%leading to the modification of jet fragmentation. 
In the present analysis, events generated from the AMPT model for Pb-Pb collisions at $\sqrt{s}$ = 2.76 TeV
have been analyzed to calculate inclusive p/$\pi$ ratio and di-hadron correlation functions  
between two sets of particles classified as {\it trigger} and {\it associated}. The p${_T}$ ranges of trigger and 
associated particles are 1.8  $<p{_T}<$ 3.0 GeV/$\it{c}$ and 1.0 $<p{_T}<$1.8 GeV/$\it{c}$ respectively and the pseudo-rapidity 
range of all particles has been restricted within -1$<\eta<$1. 
The {\it trigger} p$_T$ range has been chosen in such a way that it contains the region 
where  p/$\pi$  ratio reaches its maximum. A Two dimensional (2D) correlation function has been obtained as a function 
of the difference in 
azimuthal angle $\Delta\phi$ = $\bf\phi_{trigger}$ -$\bf\phi_{associated}$ and pseudo-rapidity $\Delta\eta$
= $\bf\eta_{trigger}$ -$\bf\eta_{associated}$. The per trigger yield of the associated particles in  $\Delta\eta$
and  $\Delta\phi$ has been defined as 
$\frac{dN_{same}}{N_{trigger}d\Delta\eta d\Delta\phi}$ where $N_{same}$ is the number of particles associated to 
the triggers particles ($N_{trigger}$) on event by event basis. 
%Both the p/$\pi$ ratio and correlation analysis have been performed dividing the entire minimum bias events
%in five centrality bins.

\begin{figure}[htbp]
%\begin{center}
\centering
\includegraphics[scale=0.5,keepaspectratio]{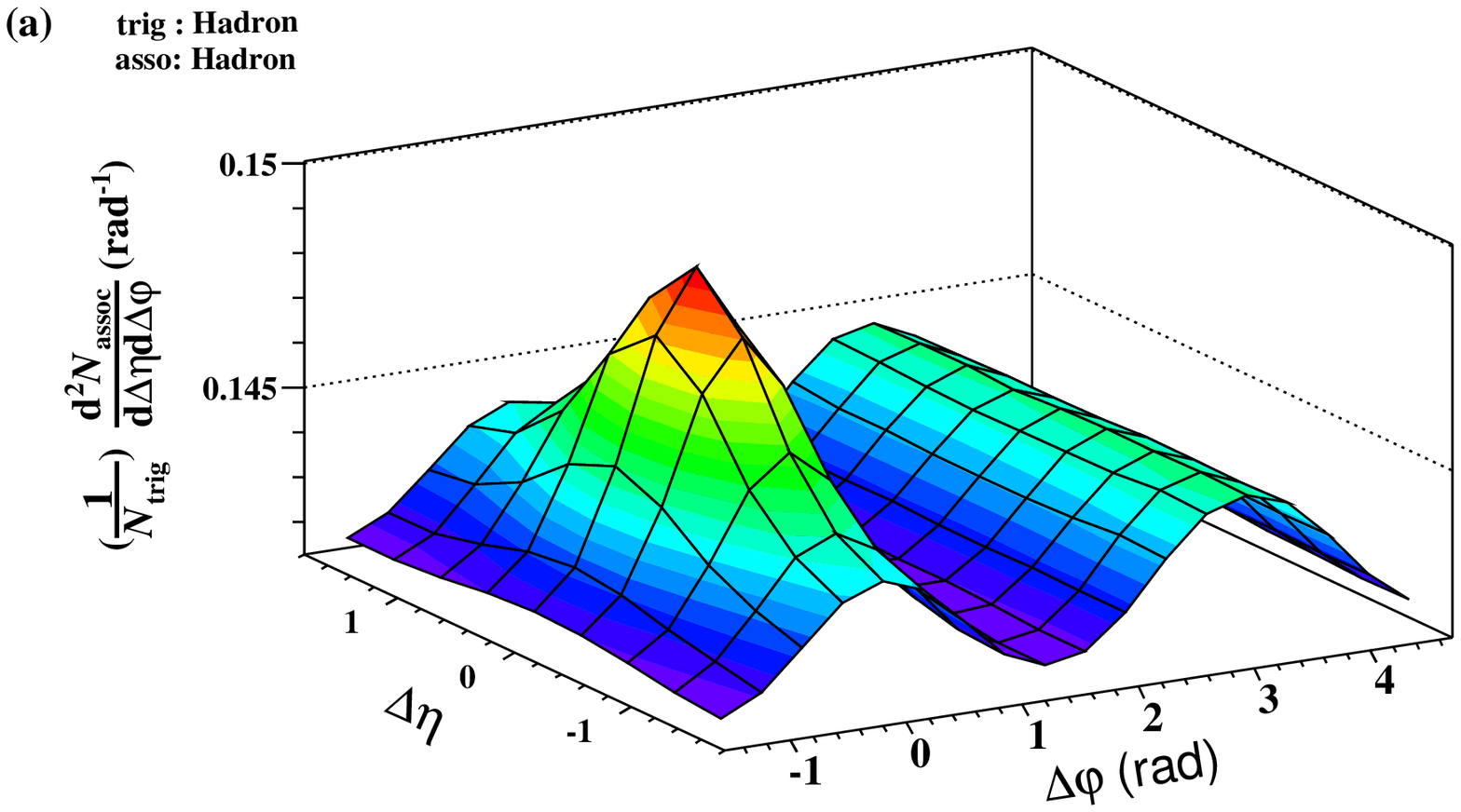}
\includegraphics[scale=0.5,keepaspectratio]{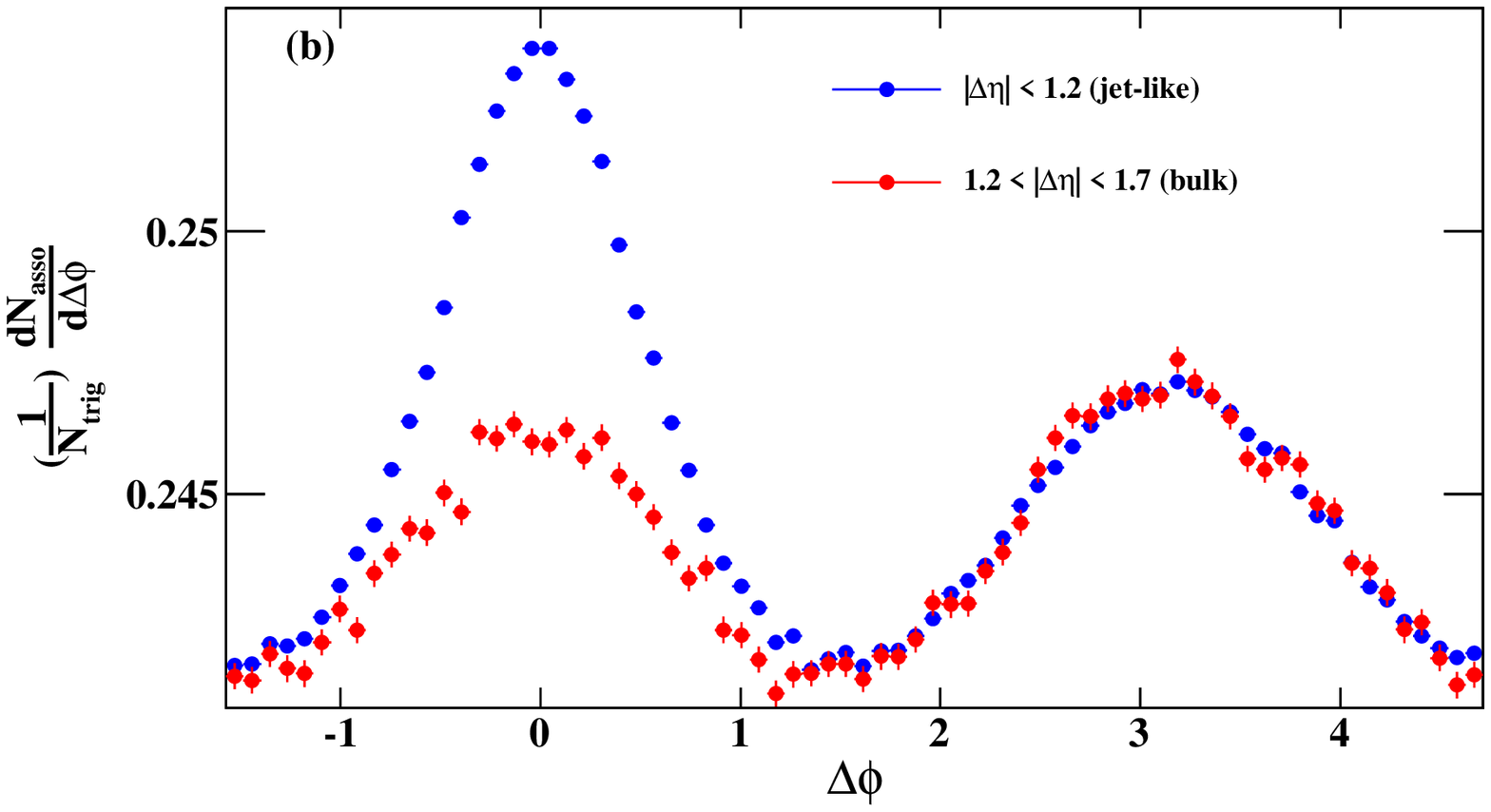}
\includegraphics[scale=0.5,keepaspectratio]{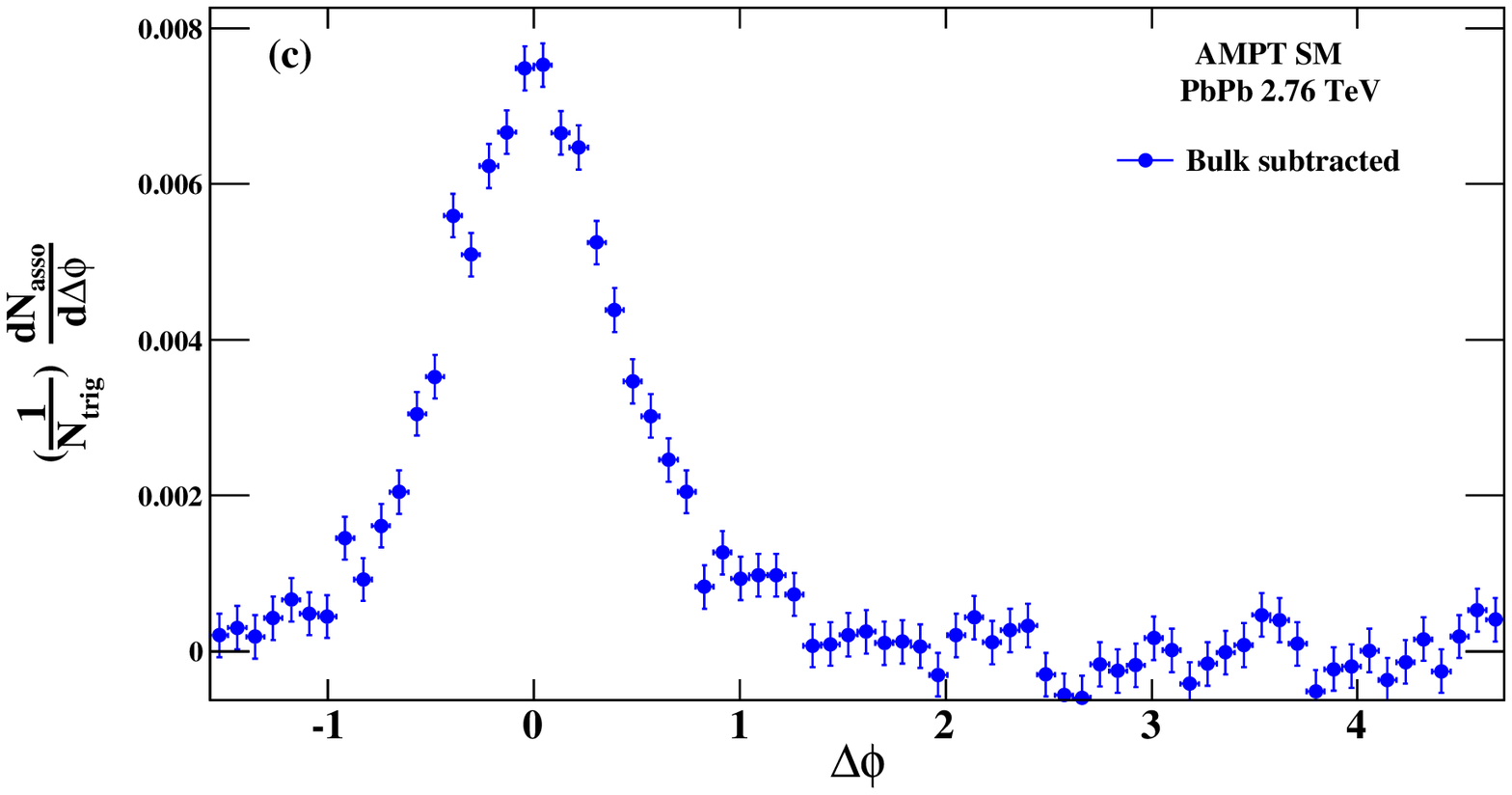}

\caption{[Color online] (a) The 2D correlation function taking unidentified charged hadrons as both trigger and associated particles.
Transverse momentum of trigger and associated particles are mentioned in the text. (b) $\Delta\phi$ projection of the 2D function for different
$\Delta\eta$ regions as indicated in the plot. (c) Near side jet-like region after bulk subtraction}
\label{2d-hadron}
%\end{center}
\end{figure}

The correlation function introduced above has been corrected for finite acceptance of trigger and associated particles. 
The acceptance corrected 2D correlation function has been obtained by dividing the raw correlations by a
correction factor represented as
$B(\Delta\eta)$ = 1 - $|\Delta\eta|$/(2.$\eta_{max}$) \cite{ALCPLB741_2015}. The correction factor has a 
triangular shape arising out of the limited acceptance in pseudo-rapidity. Uniform  2$\pi$ acceptance in azimuth ensures 
that no correction is required on $\Delta\phi$. Fig.~\ref{2d-hadron}(a) shows a corrected 2D correlation function 
for unidentified particles containing
a near side jet-like peak sitting over a flow modulated background. 

To obtain the near side jet like yield, the acceptance corrected correlation structure 
is projected on to the $\Delta\phi$ axis for $|\Delta\eta|$ $<$ 1.2.
The particles from jet fragments are most likely to be confined in a small angular region provided the width does not get 
broadened with centrality.
To isolate the contribution for near side jet-like correlations, we need to subtract
the modulation in $\Delta\phi$ arising out of the correlation with the event plane as represented by 
$v_2$,$v_3$ or higher harmonics. Flow coefficients can be extracted by fitting
the $\Delta\phi$ projection of the bulk region (large $\Delta\eta$) with $1 + 2\Sigma_{1}^{n}v_{n}^{trig}v_{n}^{asso}\rm Cos(n\Delta\phi)$ where
$v_{n}^{trig},v_{n}^{asso}$ represent the magnitude of n$^{\rm th}$ harmonic of flow coefficients for the trigger and associated particles
respectively. The Background lying beneath the jet-like peak is modulated by flow correlations dominated by elliptic flow ($v_{2}$).
We have checked that contributions from higher order flow harmonics ($v_{3}, v_{4}$) are insignificant. 
In the present analysis instead of calculating
different orders of flow harmonics and subtracting separately, we have subtracted 
the projected $\Delta\phi$ distributions at larger $\Delta\eta$ region (1.2$< |\Delta\eta| < $1.7) from the short-range region. 
The bulk subtraction by the $\eta$-gap method as stated above assumes that the correlations other than
jet-like are $\eta$  independent \cite{ALCArXiv_2013}.  
1D $\Delta\phi$ correlation functions for the regions $|\Delta\eta| <$ 1.2 and 1.2$< |\Delta\eta| < $1.7 
are shown in Fig.~\ref{2d-hadron}(b) and result from the difference of these two distribution has been ploted in Fig.~\ref{2d-hadron}(c).
The near side peak centered around  $\Delta\phi$ = 0 mainly represents jet-like correlations and the
strength of the correlation (per trigger yield) has been calculated integrating the $\Delta\phi$  distribution
over a range of $|\Delta\phi|$  $< \pi/2$.

\section{Results}

\begin{figure}[htbp]
\centering

\includegraphics[scale=0.45,keepaspectratio]{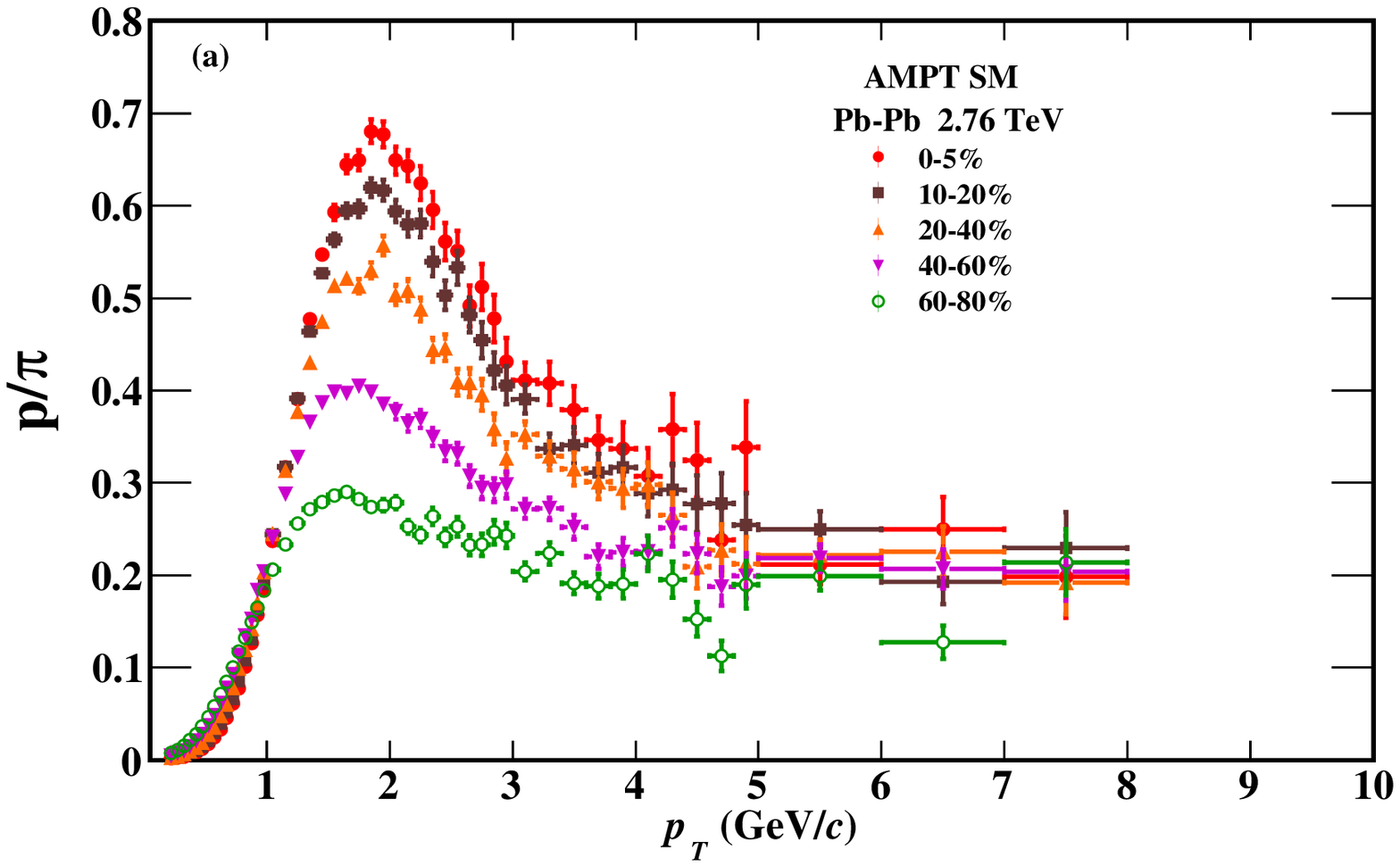}
\includegraphics[scale=0.45,keepaspectratio]{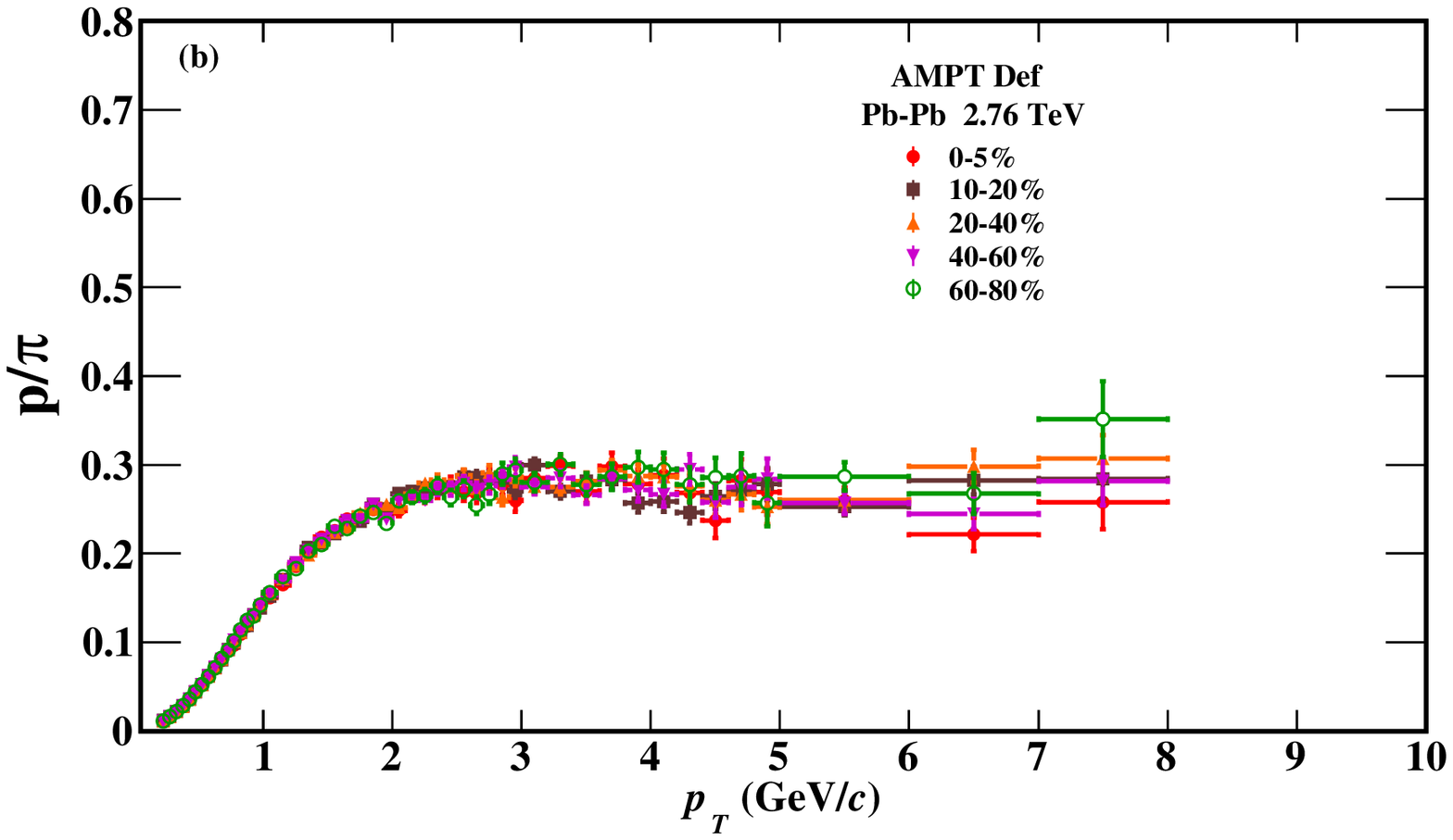}
\caption{[Color online] The ratio of the yields of proton to pion from two versions of AMPT in Pb-Pb collisions at $\sqrt{s}$ = 2.76 TeV. 
The ratio in SM version (a) shows a clear peak around 2 GeV/$\it{c}$ as opposed to the default version (b) which does not show any such peak.}
\label{ptopi}
\end{figure}

As a first step to test the features of coalescence in the SM version of AMPT, we have compared the $p_{T}$ dependence 
of the p/$\pi$ ratio from both the versions of AMPT in Fig.~\ref{ptopi}. We find a clear centrality dependence
in p/$\pi$ enhancement from the SM version of AMPT. %from peripheral to central events.
Enhancement is found to reach maximum in 0-5$\%$ most central collisions at $\approx$ 2 GeV/$\it{c}$.
In contrast, the default version shows an initial rise followed by a flat distribution of the ratio. 

\begin{figure}[htb!]
%\begin{center}
%\centering

\includegraphics[scale=0.48,keepaspectratio]{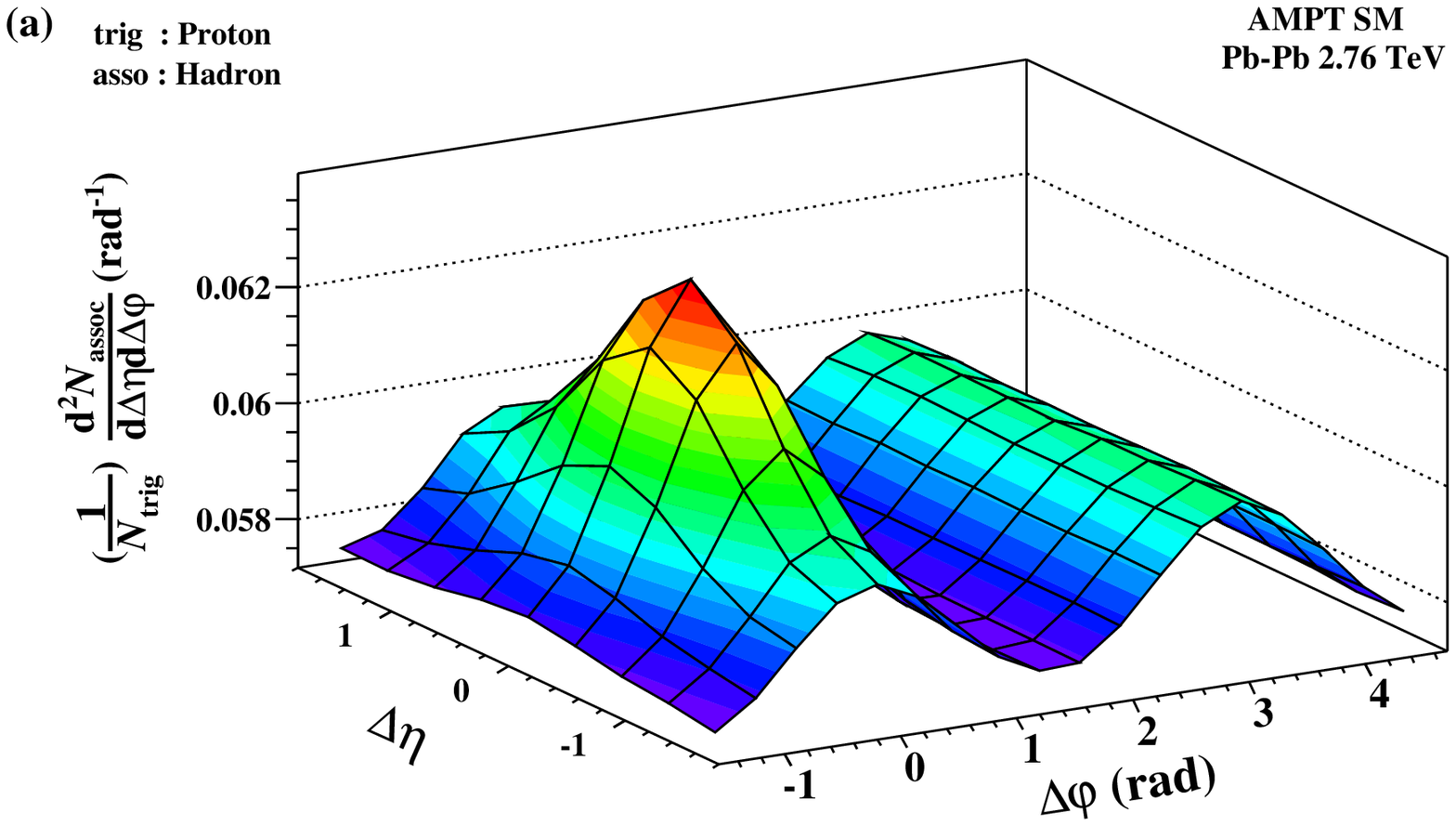}
\includegraphics[scale=0.48,keepaspectratio]{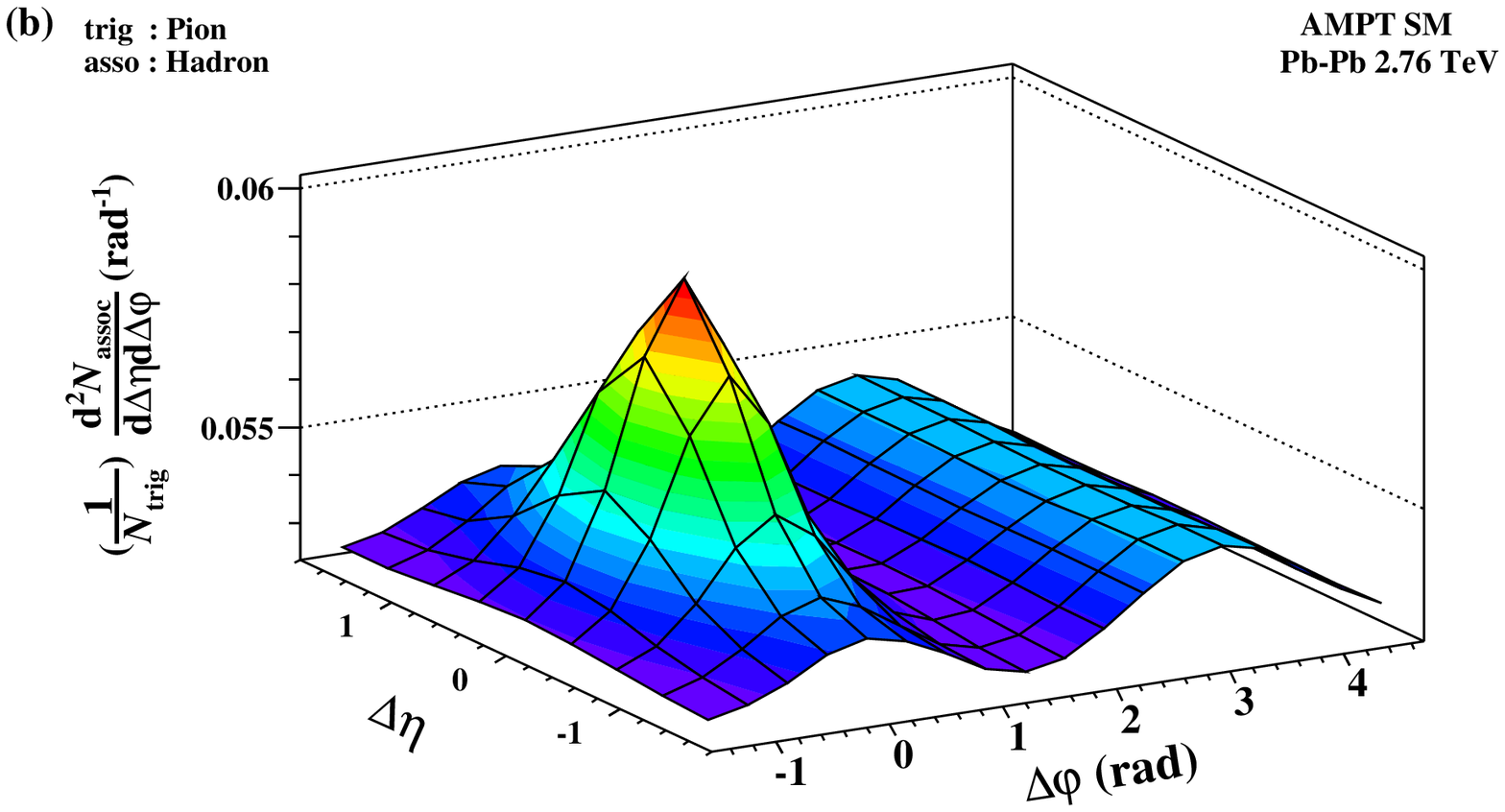}
\caption{[Color online] $\Delta\eta$-$\Delta\phi$ correlation function in 40-60 $\%$ centrality 
of Pb-Pb collisions at $\sqrt{s}$ = 2.76 TeV from AMPT SM with proton (a) and pion (b) as trigger
particles. }
%\end{center}
\label{corr-p-pi}
\end{figure}

Having established the best-known feature of the coalescence in AMPT SM model, we obtain two-particle correlations
taking leading hadrons as $\pi$ and $p/\bar{p}$  in the region where p/$\pi$ excess has been observed 
( 1.8 $\leq$  $p_{T}$ $\leq$ 3.0 GeV/$\it{c}$).
We have extracted the two-particle correlation in $\Delta\eta$-$\Delta\phi$  for 5 centrality classes selecting
trigger and associated particles in the range 1.8 $\leq$  $p_{T}$ $\leq$ 3.0 GeV/$\it{c}$ and 1.0 $\leq$  $p_{T}$ $\leq$ 1.8 GeV/$\it{c}$
respectively. Fig. ~\ref{corr-p-pi} shows the 2D $\Delta\eta$-$\Delta\phi$ correlations 
for (anti-)proton and pion as triggers. 
The per-trigger correlation functions show features typical to the presence of several effects like jet-peaks, 
harmonic coefficients among others \cite{GLMAPLB641_2006}. Near-side jet-like yield associated 
with pion and (anti-)proton triggered correlation is calculated from the $\Delta\phi$ projection of background
subtracted correlation function in the region $|\Delta\phi|$ $\leq$ $\pi/2$.
Details of background determination and subtraction have been discussed in the previous section.

\begin{figure}[htb]
\begin{center}
\includegraphics[scale=0.4,keepaspectratio]{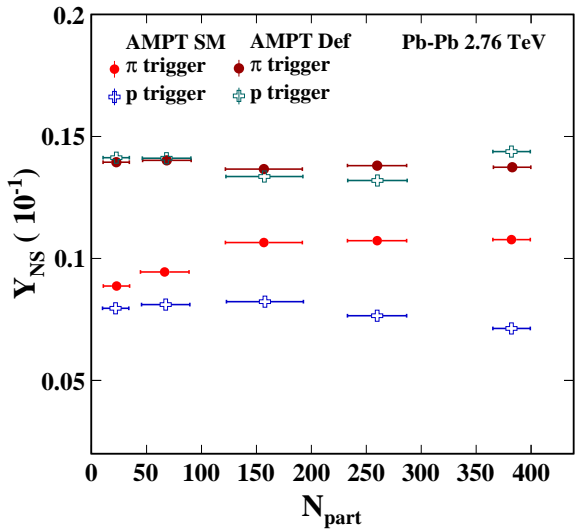}

\caption{[Color online] Centrality dependence of the near side yield from the background subtracted correlation 
function. The yields from default version of AMPT have been multiplied by 1.5 for better visualization.
 }
\label{yield-centrality-AMPTSM-Def}
\end{center}
\end{figure}

\begin{figure}[htb]
%\begin{center}

\includegraphics[scale=0.4,keepaspectratio]{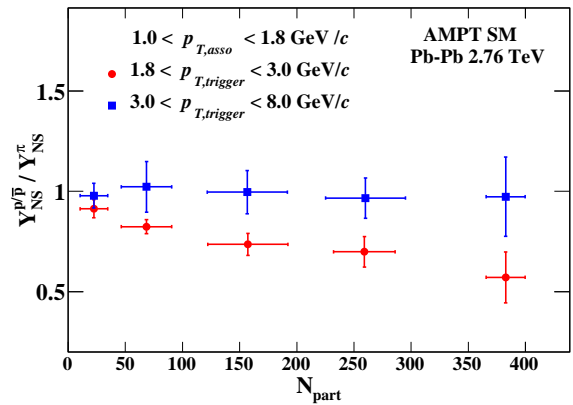}
\caption{[Color online] Ratio of proton trigger yield to pion trigger yield (Y$^{p}/$Y$^{\pi}$) at high and intermediate $p_{T,trigger}$ regions
as indicated in the figure}
\label{yield-centrality-DiffPt}
%\end{center}
\end{figure}

\begin{figure}[htb]
%\begin{center}

\includegraphics[scale=0.45,keepaspectratio]{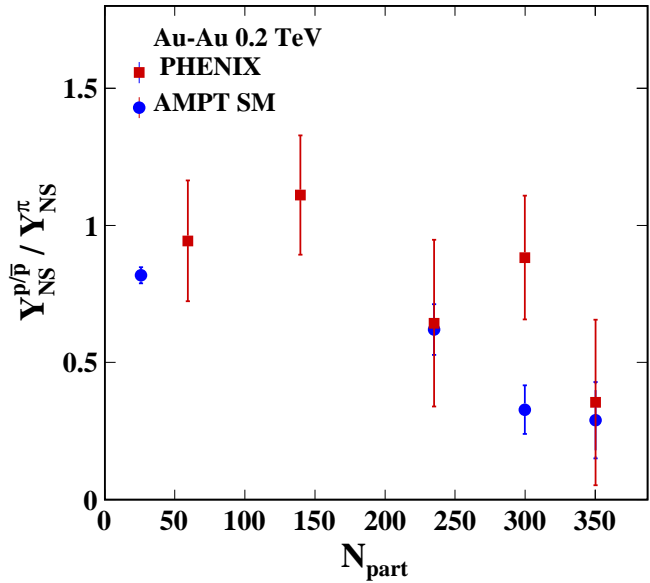}
\caption{[Color online] Ratio of proton trigger yield to pion trigger yield (Y$^{p}/$Y$^{\pi}$) for Au+Au collisions at $\sqrt(s)$ = 200 AGeV, results from PHENIX (\cite{PHCPRC71_2004} superposed.}
\label{ratio-rhic}
%\end{center}
\end{figure}

Fig.~\ref{yield-centrality-AMPTSM-Def}
shows the near-side per trigger yields as a function of centrality characterized by the number of participating nucleons (N$_{part}$) 
from the default and the SM versions. The values from the default version are multiplied by 1.5 for sake of visibility.
Yields in the default version are found to be independent of particle species and centrality. In contrast, 
in AMPT SM, pion triggered
yields are systematically higher than yields from (anti-)proton triggers over the entire centrality classes.
Interestingly both pion and proton triggered near-side yields exhibit initial rise with centrailty till N$_{part}$ $\textless$ 200.
Beyond that, per trigger yield for pion seems to attain saturation but corresponding yields for (anti-)protons undergo suppression.
In Fig.~\ref{yield-centrality-DiffPt} the ratio of the yields associated with (anti-)proton and pion triggers (Y$^{p/\bar{p}}$/Y$^{\pi}$) as a function N$_{part}$ 
have been presented for two different transverse momentum regions. In high p$_{T}$ region (3.0 $\leq$  $p_{T}$ $\leq$ 8.0 GeV/$\it{c}$),
ratios of yields are consistent with unity and independent of centrality. However, in the p$_{T}$ region 1.8 $\leq$  $p_{T}$ $\leq$ 3.0 GeV/$\it{c}$
, ratio dips showing the anticipated dilution. Similar analysis on events
generated by the SM version of AMPT for Au-Au collisions at 200 GeV and comparison with results from correlation measurements by the 
PHENIX Collaboration is represented in Fig.~\ref{ratio-rhic} It is clearly seen that the model explains dilution trend of the data qualitatively.

\section{Discussion}
 We have  measured per-trigger yield of jet-like correlations associated with pion and proton triggers
at mid-rapidity over a wide range of centrality in Pb-Pb and Au-Au collisions at $\sqrt s_{NN}$ = 2760 GeV and 200 GeV respectively,
in the momentum range where baryons are generated in excess of mesons. We have observed a significant enhancement in the jet-like
yield associated with leading pions compared to protons. In central A-A collisions, pion trigger yield is much higher than peripheral,
while the proton trigger yield exhibits a suppression. The relative enhancement in pion triggered yield could be due to
the energy dissipation of minijet partons and its re-distribution via parton cascade resulting copious production
of softer hadrons aligned to the jet-direction.

However, suppression in proton triggered yield may be attributed to the combined effect of competing processes that involve parton energy loss
and quark recombination. If protons are produced dominantly from the recombination of thermal quarks, supperssion in proton triggered
yield could be naturally expected since hadrons created by recombination of thermal partons are unlikely to have correlated partners
in small angular region. This would cause a suppression of proton trigger correlation as function of centrality as baryon generation 
at intermediate $p_{T}$ range is enhanced due to larger contribution thermal quark recombination from peripheral to central collisions.

The ratio of yields in Fig.5 shows a clear dilution in proton triggered correlation from peripheral to central events when trigger
particles are chosen from the $p_{T}$ region where inclusive p/$\pi$ ratio has shown enhancement, but no such effect has been observed
when trigger particles are chosen from higher $p_{T}$ region indicating that contributions from thermal recombination
falls-off rapidily at larger $p_{T}$.

It is interesting to note that jet-like yield calculated from the default version of AMPT has no or negligible dependence on
the choice of trigger species and almost remain unchanged with centrality. A possible reason could be that the initial
partonic density in default operation is much less than that in SM version as strings are kept intact.Thus the minijet partons during
partonic evolution suffers less interaction resulting in negligible energy dissipation. Lack of any significant energy loss
may possibly lead to no additional increase in jet-like yield.

Our study therefore indicates that the difference in jet yield of baryon-hadron and meson-hadron correlation
is an effect of competition between jet-medium interplay and dilution of jet-like yield due to quark recombination.
Comparision of our result with data would interesting as inelastic processes of energy loss are still missing in AMPT.

\section*{Acknowledgements} 
We would like to thank Dr. Partha Pratim Bhaduri and Dr. Prithwish Tribedy of VECC Kolkata for critically reading
the manuscript and suggesting necessary improvements. Thanks to VECC grid computing team for their 
constant effort to keep the facility running and helping in AMPT data generation.

\end{document}